\documentclass{article}

\usepackage{arxiv}
\usepackage{array}
\usepackage[utf8]{inputenc} % allow utf-8 input
\usepackage[T1]{fontenc}    % use 8-bit T1 fonts
\usepackage{hyperref}       % hyperlinks
\usepackage{url}            % simple URL typesetting
\usepackage{booktabs}       % professional-quality tables
\usepackage{amsfonts}       % blackboard math symbols
\usepackage{nicefrac}       % compact symbols for 1/2, etc.
\usepackage{microtype}      % microtypography
\usepackage{lipsum}		% Can be removed after putting your text content
\usepackage{graphicx}
\graphicspath{ {./fig/} }
\usepackage[square,numbers]{natbib}
\usepackage{multirow}
\usepackage{doi}
\usepackage{nomencl}
\usepackage[ruled]{algorithm2e}
\usepackage{float}
\usepackage[table,xcdraw]{xcolor} % 用于表格背景色
\usepackage{tcolorbox}

\title{Large Language Models for Combinatorial Optimization of Design Structure Matrix}

\date{A Preprint, Version: Nov 19, 2024} 					% Or removing it

\author{ 
{
\hspace{1mm}Shuo Jiang}\thanks{Comments are welcome: \texttt{shuojiangcn@gmail.com}} \\
	Department of Systems Engineering\\
	City University of Hong Kong\\
	83 Tat Chee Ave, Kowloon Tong, Hong Kong\\
	\texttt{shuo.jiang@cityu.edu.hk} \\
	\And
	{\hspace{1mm}Min Xie} \\
        Department of Systems Engineering\\
	City University of Hong Kong\\
	83 Tat Chee Ave, Kowloon Tong, Hong Kong\\
	\texttt{xiemin@cityu.edu.hk} \\
        \And
	{\hspace{1mm}Jianxi Luo} \\
        Department of Systems Engineering\\
	City University of Hong Kong\\
	83 Tat Chee Ave, Kowloon Tong, Hong Kong\\
	\texttt{jianxi.luo@cityu.edu.hk} \\
}

\renewcommand{\headeright}{Large Language Models for Combinatorial Optimization of DSM}
\renewcommand{\undertitle}{}
\renewcommand{\shorttitle}{}

\begin{document}
\maketitle

\begin{abstract}
	Combinatorial optimization (CO) is essential for improving efficiency and performance in engineering applications. As complexity increases with larger problem sizes and more intricate dependencies, identifying the optimal solution become challenging. When it comes to real-world engineering problems, algorithms based on pure mathematical reasoning are limited and incapable to capture the contextual nuances necessary for optimization. This study explores the potential of Large Language Models (LLMs) in solving engineering CO problems by leveraging their reasoning power and contextual knowledge. We propose a novel LLM-based framework that integrates network topology and domain knowledge to optimize the sequencing of Design Structure Matrix (DSM)—a common CO problem. Our experiments on various DSM cases demonstrate that the proposed method achieves faster convergence and higher solution quality than benchmark methods. Moreover, results show that incorporating contextual domain knowledge significantly improves performance despite the choice of LLMs. These findings highlight the potential of LLMs to address complex CO problems by combining semantic and mathematical reasoning. This approach paves the way for a new paradigm in real-world engineering combinatorial optimization.
\end{abstract}


\section{Introduction}
\label{sec1}

Combinatorial optimization (CO) problems are ubiquitous across fields, where finding an optimal solution from a finite set often drives improvements in efficiency, cost, and performance \cite{korte2011combinatorial}. For instance, applications such as DNA barcoding and DNA assembly in synthetic biology \cite{naseri2020synthetic}, as well as job scheduling in manufacturing \cite{xidias2019motion}, rely heavily on effective CO solutions. However, due to their NP-hard nature, these problems present substantial challenges, especially as complexity increases with larger problem sizes and more intricate dependencies. Traditionally, CO problems in engineering are usually approached through the following process: the problem is first modelled mathematically, then solved using specific algorithms or heuristics, and finally interpreted within the context of practical engineering \cite{pinedo2012scheduling}. This separation of problem-solving and interpretation stages is limited and incapable to capture the contextual nuances necessary for optimization of real-world problems.

Recent advancements in Large Language Models (LLMs) have demonstrated their powerful capabilities in natural language generation, semantic understanding, instruction following, and complex reasoning \cite{wei2022emergent, chang2024evaluation}. Furthermore, pioneering studies have shown that LLMs can be used for continuous and concrete optimization \cite{romera2024math,yang2024optimizers,liu2024evolutionary}. For instance, researchers from DeepMind utilized LLMs as optimizers and evaluated their effectiveness on classic CO problems \cite{yang2024optimizers}, such as Traveling Salesman Problems (TSP). Additionally, prior studies also highlight that LLMs possess extensive domain knowledge pretrained across a wide range of engineering-related data, which enhances their applicability in engineering fields \cite{zhu2023transformers,makatura2024design,jiang2024autotriz,mei2024replanvlm}. Therefore, the ability of LLMs to combine mathematical and semantic reasoning, along with their possession of extensive knowledge, motivated us to explore their potential for solving engineering CO problems while integrating contextual domain knowledge relevant to their network typology. Our hypotheses are: (1) LLMs can be effectively applied to solve CO problems in engineering, and (2) incorporating contextual domain knowledge can further enhance LLM performance by supporting mathematical reasoning with semantic insights. This paradigm, which leverages both semantic and mathematical reasoning, introduces a novel approach to combinatorial optimization that traditional pure mathematical methods cannot achieve for empirical problems. On this basis, we propose a novel LLM-based framework that integrates both network topology and domain context into the optimization process.

\begin{figure}[H]
	\centering
	\includegraphics[width=15cm]{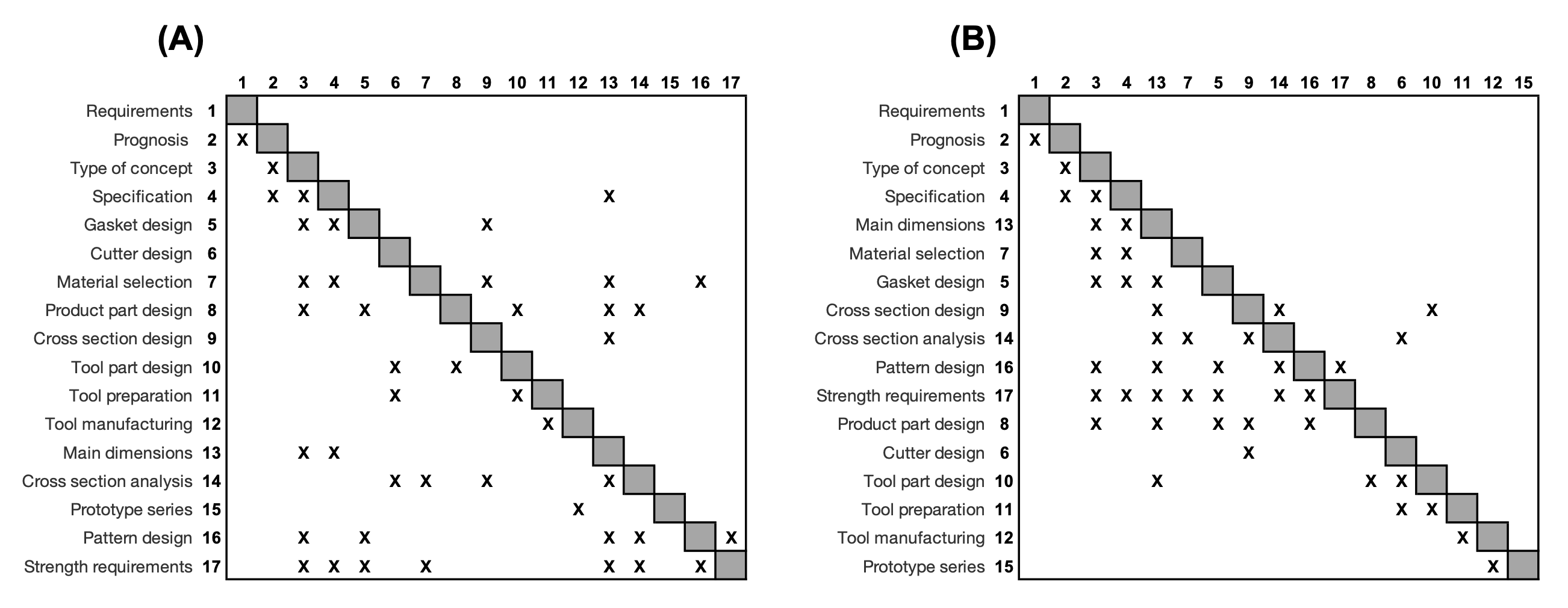}
	\caption{Illustration of a Design Activity DSM: (A) Pre-Sequencing; (B) Post-Sequencing}
	\label{fig:fig1}
\end{figure}

To evaluate our proposed method, we focus on the Design Structure Matrix (DSM) sequencing task, as an example of CO problems. DSM is a modelling tool in engineering design, which represents dependency relationships among tasks or components within a system \cite{steward1981design}. Reordering the node sequence of DSMs can significantly reduce feedback loops and improve modularization \cite{choi2011scheduling,eppinger2012dsm}. The DSM sequencing is also an NP-hard problem, and traditional methods typically approach it using heuristics-based algorithms \cite{eppinger1994tasks,qian2011dsm}. Figure \ref{fig:fig1} illustrates a design activity DSM before and after sequencing \cite{amen1999matching}. In this paper, we conduct extensive experiments on various DSM cases to demonstrate that our LLM-based method achieves better convergence speed and solution quality compared to benchmark methods. Notably, results show that incorporating contextual domain knowledge significantly enhances the performance despite the choice of backbone LLMs.

\section{Methodology}
\label{sec:sec2}

\begin{figure}[H]
	\centering
	\includegraphics[width=15cm]{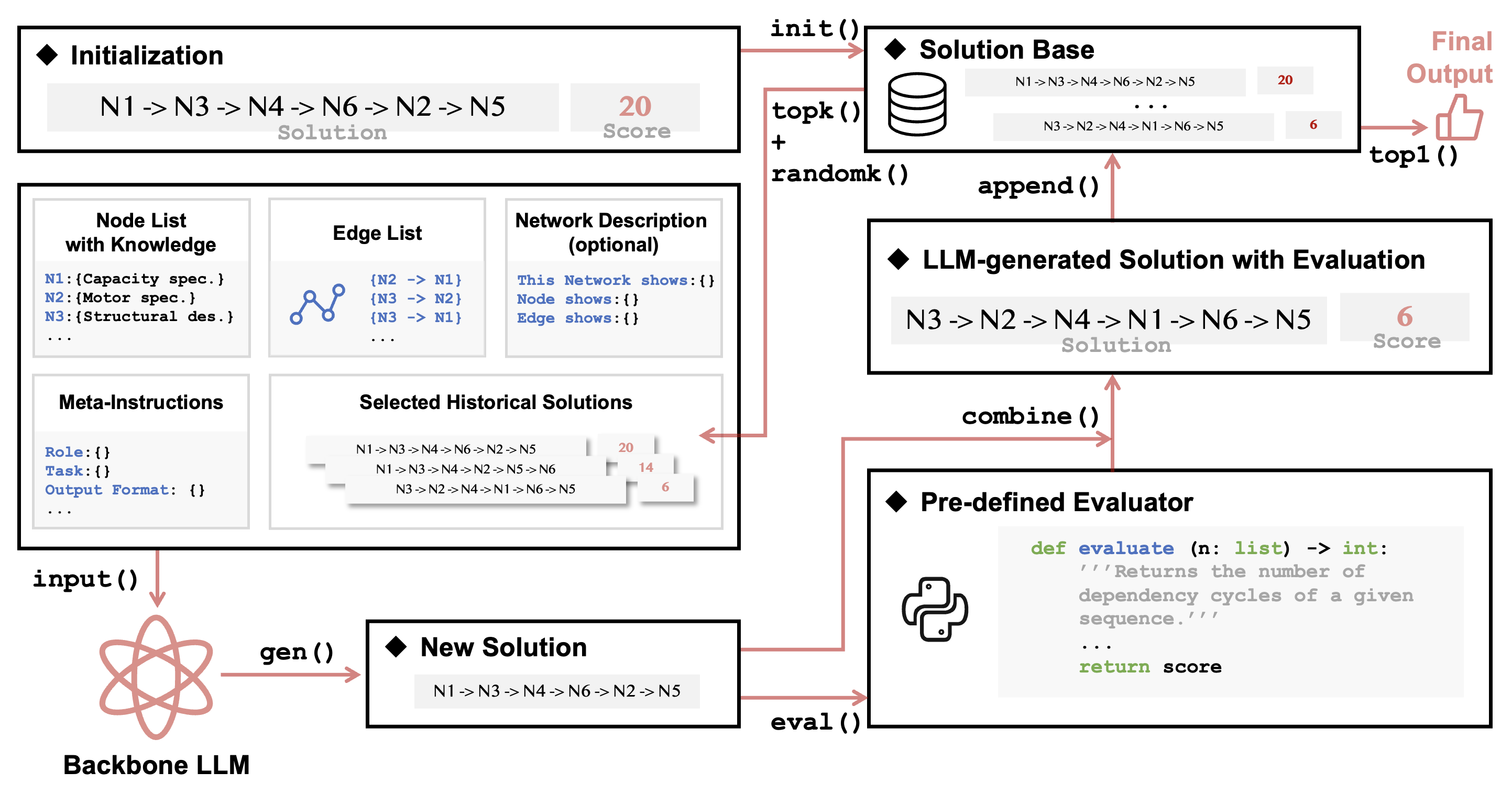}
	\caption{Overview of the Proposed Framework}
	\label{fig:fig2}
\end{figure}

In this section, we introduce our proposed LLM-based framework for solving CO problems. The framework is designed to harness the generative and reasoning powers of LLMs in combination with domain knowledge and objective evaluation. The framework begins with the initialization of a solution randomly sampled in the total solution space. Each solution is evaluated based on predefined criteria by an evaluator, which quantifies the quality of a solution. Using this evaluation, the framework iteratively updates the solution base through few-shot learning and suggesting new candidates, guided by crafted prompts that include both network information in mathematical form and domain knowledge in natural language description. The newly generated solutions are appended into the solution base, together with their evaluation results. When the iteration time is reached, the solution base returns the best one as the final output. In following, we focus on DSM sequencing as a common CO problem to illustrate the pipeline. The framework is depicted in Figure \ref{fig:fig2}.

\textbf{Initialization and solution sampling}

The Solution Base serves as an essential module to (1) enable storage of explored solutions and their evaluation results, (2) provide historical solutions to the backend LLM for few-shot learning, and (3) return the top-performing solution when iteration ends.

The initialization involves generating an initial solution that is randomly sampled from the entire solution space. In the DSM sequencing task, a solution represents a complete and non-repetitive sequence of nodes (see Figure \ref{fig:fig2}). This initial solution is then evaluated and added to the Solution Base for future use. In subsequent iterations, we design a sampling rule that selects $K_p$ top-performance solutions and randomly samples $K_q$ solutions from the remaining $K_n-K_p$ solutions to form a solution set, where $K_n$ is the total number of solutions in the Solution Base. $K_p$ and $K_q$ are adjustable parameters. The obtained solution set is further refined and crafted into prompts.

\textbf{LLM-driven optimization using network information and domain knowledge}

In each iteration of the optimization process, we prompt backend LLM with the following information: \textbf{(i) Typology Information}: These two elements complete the mathematical description of a DSM. It is noteworthy that there are multiple equivalent representations for describing a network mathematically, such as an edge list, a dependency relationship list according to node sequence, or an adjacency matrix. In this research, we choose the edge list as the representation of the network's topology and shuffle all edges to avoid any possible bias. \textbf{(ii) Contextual Domain Knowledge}: This includes the name of each node and an overall description of the network, which conveys the domain knowledge underlying the DSM’s mathematical structure to the LLM. For instance, in an activity DSM, each node represents the name of an activity in the entire design process. \textbf{(iii) Meta-instructions}: We adopt some frequently used prompt engineering strategies \cite{zhao2023survey}, including role-playing, task specification, and output format specification. These strategies allow the LLM to follow the guidance to perform reasoning and generate solutions in a specific format. \textbf{(iv) Selected historical solutions}: As described in the last section, we obtain a set of up to $K_p+K_q$ solutions through sampling from the Solution Base for the LLM to use in few-shot learning.

Once receiving inputs described above, the backend LLM combines the network topology information with domain knowledge to infer and suggest new solutions. The generated solution must pass the checker, which verifies that all nodes are present exactly once in the sequence. Once validated, the solution is then evaluated and appended to the Solution Base. The detailed input prompts are included in Appendix 1.

\textbf{Evaluation of DSM sequencing solutions}

The evaluator is used to quantify each newly generated solution. For the DSM sequencing task, the goal is to reorder the rows and columns of the DSM to minimize feedback loops. To achieve this objective, the evaluator calculates the number of backward dependencies in the corresponding sequence.

We formally describe this problem as follows: Given a sequence $s$, we can use the edge list of the network to obtain an $n \times n$ asymmetric adjacency matrix $A$, representing a directed network of $n$ nodes, where $a_{ij}$ is a typical binary entry. Specifically, $a_{ij} = 1$ represents that node $i$ depends on node $j$. The task is to reorder the $n$ correlated rows and columns to minimize the number of entries above the main diagonal in the corresponding adjacency matrix. On this basis, the objective function can be formally expressed as: $\min_{s} \sum_{i=1}^{n} \sum_{j=i+1}^{n} a_{ij}$, where $i<j$.

For each sequence generated by the LLM (including the initial randomly created one), we calculate its evaluation score using the formula above. In this task, a lower score indicates a higher quality solution.

\section{Experiments}
\label{sec3}
\subsection{Data}

We collected four DSM cases for our experiments, which can be categorized into two types \cite{browning2001dsm}: (1) Activity-based DSMs, which represent the input-output relationships between different tasks or activities within a project; and (2) Parameter-based DSMs, which illustrate the relationships among design parameters of a product. 

The DSM of the Unmanned Combat Aerial Vehicle (UCAV) includes 12 conceptual design activities conducted at Boeing \cite{browning1998modeling}. The DSM of the Microfilm Cartridge was derived from Kodak's Cheetah project including 13 major tasks \cite{ulrich2016product}. These two activity-based DSMs were constructed based on interviews with relevant engineers, followed by review, verification, and calibration to ensure accuracy. For the parameter-based DSMs, the Heat Exchanger \cite{amen1999matching} and the Automobile Brake System \cite{black1990method}, researchers first identified the key components from the product and then interviewed the corresponding designers to define design parameters and establish precedence relationships. The Heat Exchanger DSM contains 17 components related to core thermal exchange elements, while the Brake System DSM includes 14 main parameters covering braking mechanisms and their dependencies.

From the referential documents, we extracted: (1) the name of each node, (2) the edge list obtained from the adjacency matrix, and (3) the overall description of each network. All data were kept consistent with the original references. The specific data formats are shown in Appendix 1. The characteristics of four DSMs are summarized in Table \ref{tab:table1}. In general, the node count (\textbf{N}) of DSMs ranges from 12 to 17, and the edge count (\textbf{E}) varies between 32 and 47. The network diameter, representing the longest shortest path between any two nodes, spans from 2 to 7. We also present measures such as network density and clustering coefficient to highlight the complexity. For instance, the UCAV DSM has a high network density of 0.712 and a clustering coefficient of 0.773, suggesting strong interconnections, while measures of the Heat Exchanger DSM indicate a sparser network with more distinct relationships.

\begin{center}
\begin{table}[H]
	\caption{Characteristics of Four DSMs}
	\centering
	\renewcommand{\arraystretch}{1.5}
	\begin{tabular}{m{4.2cm}<{\centering}m{0.6cm}<{\centering}m{0.6cm}<{\centering}m{1.45cm}<{\centering}m{1.3cm}<{\centering}m{1.3cm}<{\centering}m{1.7cm}<{\centering}m{1.95cm}<{\centering}}
		\toprule
		\textbf{} & \textbf{N} & \textbf{E} & \textbf{Network Diameter} & \textbf{Network Density} & \textbf{Average Degree} & \textbf{Clustering Coefficient} & \textbf{Average Path Length} \\
		\midrule
		
		\multirow{2}{6cm}{\textbf{Activity-Based DSMs} \\ Unmanned Aerial Vehicle \cite{browning1998modeling} \\ Microfilm Cartridge \cite{ulrich2016product}} 
		& 12 & 47 & 2 & 0.712 & 7.833 & 0.773 & 1.288 \\
		& 13 & 41 & 3 & 0.526 & 6.308 & 0.682 & 1.577 \\
        
        \midrule
        
        \multirow{2}{6cm}{\textbf{Parameter-Based DSMs} \\ Heat Exchanger \cite{amen1999matching} \\ Automobile Brake System \cite{black1990method}} 
        & 17 & 41 & 7 & 0.302 & 4.824 & 0.457 & 2.397 \\
		& 14 & 32 & 4 & 0.352 & 4.571 & 0.414 & 1.824 \\
		
		\bottomrule
	\end{tabular}
	\label{tab:table1}
\end{table}
\end{center}

\subsection{Experiment Setup}

\textbf{Experimental setting}

In all our experiments, we ran each method 10 times with different random seeds to ensure robustness. For solution sampling, we set $K_p=5$ and $K_q=5$. For the index of each node, we use a string composed of 5 randomly generated characters (a mix of numbers and letters) to represent the node uniquely. The maximum number of iterations was set to 20. We selected \textit{Claude-3.5-Sonnet-20241022} as the backbone LLM and kept all other settings of the LLM as their default values \cite{anthropic2024claude}.

\textbf{Comparative methods for benchmarking}

We consider two types of approaches for benchmarking.

The first type is stochastic methods, which rely on probabilistic rules to explore the solution space. We chose the classic Genetic Algorithm (GA) for comparison, as it has been successfully applied to various optimization problems \cite{alhijawi2024genetic}. We implemented GA using the DEAP library \cite{fortin2012deap}. We considered three different settings for GA: exploration-focused, exploitation-focused, and balanced setting. Specific parameter settings of three variants are detailed in Appendix 2.

The second type is deterministic methods, which rank nodes based on a specific measure and then use this ranking for reordering. Since the calculation of each measure is deterministic, the resulting solution is consistent. We considered five deterministic methods for comparison, each reordering nodes based on different criterion: \textbf{(i) Out-In Degree} \cite{crofts2011brain}: Reordering based on the difference between out-degree and in-degree of each node. \textbf{(ii) Eigenvector} \cite{dietzenbacher1992interindustry}: Calculating the values of components in the Perron vector of the adjacency matrix and sorting nodes accordingly. \textbf{(iii) Walk-based (Exponential)} \cite{estrada2005subgraph}: Considering in-depth connectivity patterns by involving the power of the adjacency matrix \textbf{A}. It is calculated by: $F(\mathbf{A}) = \exp(\mathbf{A})$. \textbf{(iv) Walk-based (Resolvent)} \cite{estrada2010network}: Calculated by: $F(\mathbf{A}) = (\mathbf{I} - \delta\mathbf{A})^{-1}$, where $\delta$ represents the probability that a message will successfully traverse an edge. We set $\delta=0.025$ in experiments following the original reference. \textbf{(v) Visibility} \cite{maccormack2012duality}: Involving the visibility matrix, showing the dependencies between all system elements for all possible path lengths up to the matrix size. It is calculated by: $F(\mathbf{A}) = \sum_{k=0}^{n} \mathbf{A}^k$, where n represents the number of nodes. The resulting visibility matrix is then binarized. For methods \textbf{(iii)}, \textbf{(iv)}, and \textbf{(v)}, once the $F(\mathbf{A})$ is obtained, nodes are reordered by the sum of rows of $F(\mathbf{A})$. If nodes share the same value, they are then reordered based on the sum of columns of $F(\mathbf{A})$.

To further investigate the effectiveness of incorporating domain knowledge, we also consider a variant of our proposed method for comparison. In this variant, all settings are kept the same as in the main method, except that the contextual knowledge about nodes and the entire network is removed. Therein, only the topological information of the network is engineered into the input prompt, guiding the LLM to search for an optimal solution solely through mathematical reasoning.

\section{Results and Discussion}
\label{sec:sec4}

\begin{figure}
	\centering
	\includegraphics[width=15cm]{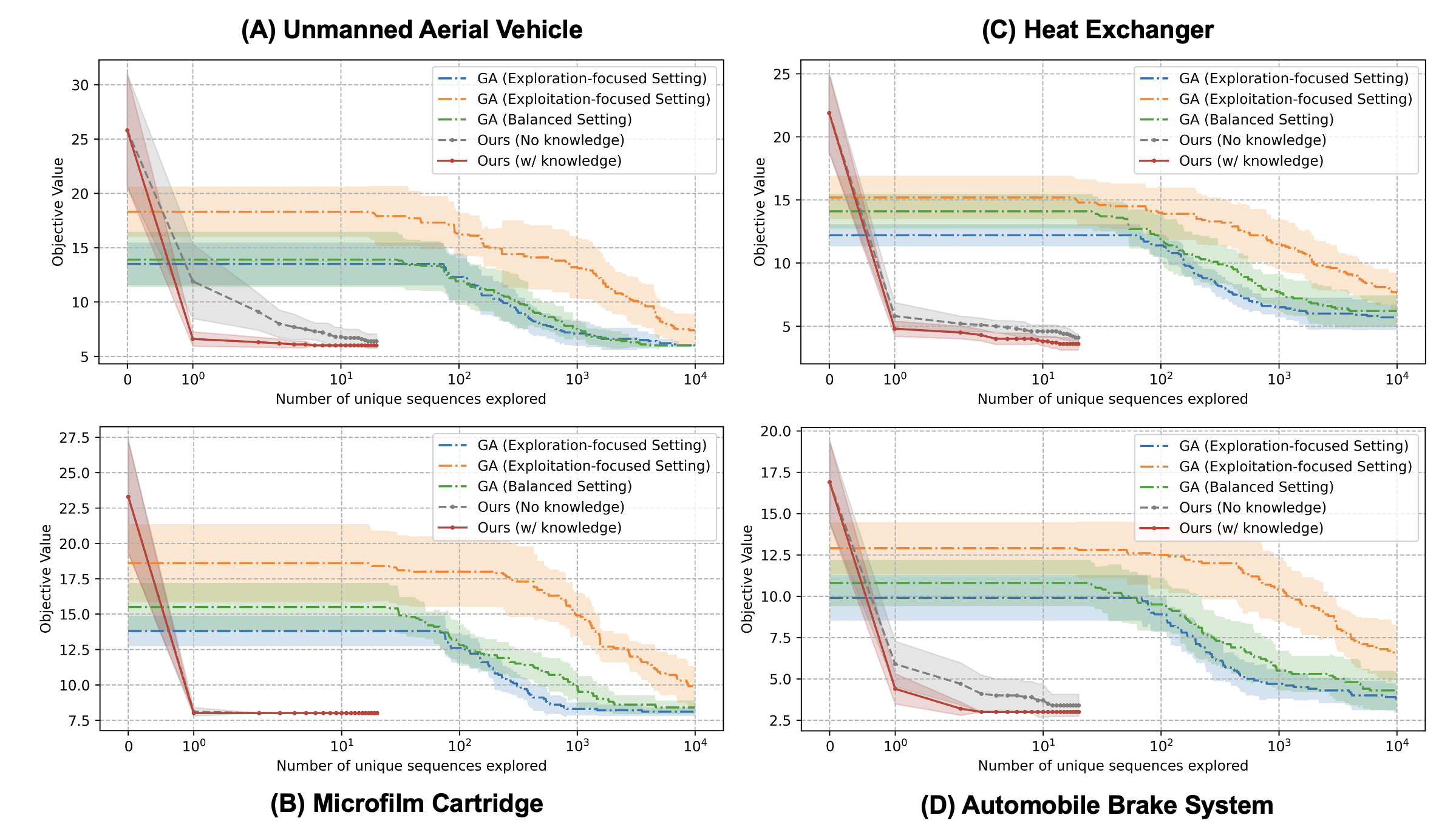}
	\caption{Comparison of Convergence Speed}
	\label{fig:fig3}
\end{figure}

We first compared the convergence speed among our LLM-based methods and the three variants of GA, as shown in Figure \ref{fig:fig3}. In GA, due to different parameter settings (e.g., population size, crossover, and mutation rates), each iteration explores multiple unique solutions. In contrast, our methods suggest only one unique solution per iteration. Therefore, we standardized the iteration time (generation) for GA as the number of unique solutions explored and used this as the measure for convergence speed. Although the maximum number of generations for each GA run was set to 2,000 (refer to Appendix 2), we truncated the comparison to the first 10,000 unique solutions explored for clearer visualization.

Compared to all three GA settings, our methods show significantly higher convergence speed, reaching lower objective values faster across all four cases. Overall, our LLM-based methods demonstrate a consistent advantage over GA, finding high-quality solutions multiple orders of magnitude faster in all cases. In each of the four cases, our methods can always identify an initial solution of good quality within the first attempt and continue optimizing the solution through subsequent iterations, approximating the optimal solution. This demonstrates that LLMs can leverage in-context learning to adapt and improve solutions efficiently.

Moreover, in Figures \ref{fig:fig3} (A), (C), and (D), we observe that incorporating domain knowledge enhances the LLM's performance in both convergence speed and solution quality, as indicated by the red line consistently lying below the grey line. This suggests that LLMs combining semantic reasoning with mathematical reasoning can solve CO problems more effectively and efficiently. It is noteworthy that in the cartridge development case (Figure \ref{fig:fig3} (B)), both LLM-based methods (with and without knowledge) found the best solution within the first step. Consequently, the two lines overlap, indicating identical performance in this case.

\begin{table}[H]
    \caption{Comparison of Solution Quality}
    \centering
    \renewcommand{\arraystretch}{1.3}
    \setlength{\tabcolsep}{6pt} % 调整列间距
    \begin{tabular}{l m{2.3cm}<{\centering} m{2.3cm}<{\centering} m{2.3cm}<{\centering} m{2.3cm}<{\centering}}
        \toprule
        \textbf{Methods} & \multicolumn{2}{c}{\textbf{Activity-Based DSMs}} & \multicolumn{2}{c}{\textbf{Parameter-Based DSMs}} \\
        \cmidrule(lr){2-3} \cmidrule(lr){4-5}
        & \textbf{Unmanned Aerial Vehicle} & \textbf{Microfilm Cartridge} & \textbf{Heat Exchanger} & \textbf{Automobile Brake System} \\
        \midrule
        \textbf{Stochastic Methods\footnotemark[1]} & & & & \\
        GA (Exploration-focused setting) & \textbf{6.0±0.0} & 8.1±0.3 & 5.7±1.0 & 3.8±0.7 \\
        GA (Exploitation-focused setting) & 7.4±1.5 & 9.9±1.4 & 7.6±1.4 & 6.6±1.7 \\
        GA (Balanced setting) & \textbf{6.0±0.0} & 8.4±0.5 & 6.2±1.2 & 4.2±1.2 \\
        \midrule
        \textbf{Deterministic Methods\footnotemark[2]} & & & & \\
        Out-In Degree \cite{crofts2011brain} & 10.0±0.0 & 12.3±0.5 & 10.3±0.6 & 5.7±0.9 \\
        Eigenvector \cite{dietzenbacher1992interindustry} & 15.0±0.0 & 13.8±0.7 & 13.1±1.7 & 11.1±1.4 \\
        Walk-based (Exponential) \cite{estrada2005subgraph} & 15.0±0.0 & 12.0±0.0 & 8.0±0.0 & 11.0±0.0 \\
        Walk-based (Resolvent) \cite{estrada2010network} & 9.0±0.0 & 12.0±0.0 & 8.0±0.0 & 11.0±0.0 \\
        Visibility \cite{maccormack2012duality} & 25.6±2.2 & 8.8±0.7 & 6.0±1.3 & \textbf{3.0±0.0} \\
        \midrule
        \textbf{LLM-driven Methods (Ours)} & & & & \\
        Single-trial with knowledge & 6.6±0.7 & \textbf{8.0±0.0} & 4.8±0.6 & 4.4±0.9 \\
        Single-trial without knowledge & 11.9±3.4 & 8.1±0.3 & 5.8±1.1 & 5.9±1.4 \\
        5-trial with knowledge & 6.1±0.3 & \textbf{8.0±0.0} & 4.0±0.4 & \textbf{3.0±0.0} \\
        5-trial without knowledge & 7.5±0.9 & \textbf{8.0±0.0} & 4.9±0.5 & 4.0±1.0 \\
        20-trial with knowledge & \textbf{6.0±0.0} & \textbf{8.0±0.0} & \textbf{3.6±0.5} & \textbf{3.0±0.0} \\
        20-trial without knowledge & 6.4±0.7 & \textbf{8.0±0.0} & 4.1±0.3 & 3.4±0.7 \\
        \bottomrule
    \end{tabular}
    \label{tab:table2}
\end{table}
\footnotetext[1]{Consistent with Figure \ref{fig:fig3}, all three GA settings show optimization performance with number of unique sequences explored = 10,000.}
\footnotetext[2]{All deterministic methods can be regarded as single-trial approaches, as they produce consistent outcomes for identical inputs across repeated executions. However, certain nodes might share the same measures. For these nodes, a random order is applied, and the final statistical evaluation results are obtained from 10 runs.}

We also compared the solution quality among different methods, as presented in Table \ref{tab:table2}. We experimented with our methods over single-trial, 5-trial, and 20-trial runs to evaluate their effectiveness. Compared to stochastic methods and deterministic methods, our methods significantly outperformed the corresponding benchmarks.

On the one hand, since all deterministic methods can be regarded as single-trial approaches (static algorithms), we first compared our single-trial results with those methods. As shown in Table \ref{tab:table2}, except for the Visibility-based method finding the optimal solution directly in one case (brake system design DSM), our methods outperformed all other deterministic methods across all cases. Even in the brake system case, our 5-trial method matched the effectiveness of the Visibility-based method.

On the other hand, when comparing with stochastic algorithms, our 20-trial method consistently found better solutions than benchmarks and achieved the optimal solution in 3 out of 4 cases. It is important to note that our experiments only tested up to 20 iterations. If we were to continue with longer iterations, the heat exchanger design case might also reach the optimal solution. Moreover, even with the 5-trial runs, our methods outperformed all three variants of GA in 3 out of 4 cases and were close in the UCAV case (6.1±0.3 vs. 6.0±0.0). This demonstrates that our LLM-based approach, leveraging contextual domain knowledge and iterative in-context learning, can effectively solve CO problems, suggesting high-quality solutions in a few attempts.

\begin{table}[H]
    \caption{Ablation on the Backbone LLM}
    \centering
    \renewcommand{\arraystretch}{1.2} % 调整行间距
    \setlength{\tabcolsep}{4pt} % 调整列间距
    \begin{tabular}{p{1.6cm}<{\centering} m{2.8cm} m{1.5cm}<{\centering} p{2.15cm}<{\centering} p{2.15cm}<{\centering} p{2.15cm}<{\centering} p{2.15cm}<{\centering}}
        \toprule
        \multirow{2}{*}{\textbf{\# of Trial}} & \multirow{2}{*}{\textbf{Backbone LLM}} & \multirow{2}{*}{\textbf{Knowledge}} 
        & \multicolumn{2}{c}{\textbf{Activity-Based DSMs}} 
        & \multicolumn{2}{c}{\textbf{Parameter-Based DSMs}} \\ 
        \cmidrule(lr){4-5} \cmidrule(lr){6-7}
        & & & \textbf{Unmanned Aerial Vehicle} & \textbf{Microfilm Cartridge} 
        & \textbf{Heat Exchanger} & \textbf{Automobile Brake System} \\
        \midrule
        \multirow{8}{*}{\textbf{1}} 
        & \multirow{2}{*}{\textit{Mixtral-7x8B}} & \cellcolor[HTML]{EFEFEF}with    & \cellcolor[HTML]{EFEFEF}12.5±2.2 & \cellcolor[HTML]{EFEFEF}19.5±5.0 & \cellcolor[HTML]{EFEFEF}16.1±2.2 & \cellcolor[HTML]{EFEFEF}12.1±3.1 \\
        &                               & without & 15.8±4.6 & 17.7±3.3 & 16.8±1.9 & 16.0±2.3 \\ 

        & \multirow{2}{*}{\textit{Llama3-70B}}  & \cellcolor[HTML]{EFEFEF}with    & \cellcolor[HTML]{EFEFEF}7.6±0.8  & \cellcolor[HTML]{EFEFEF}8.3±0.5  & \cellcolor[HTML]{EFEFEF}7.9±2.0  & \cellcolor[HTML]{EFEFEF}6.1±1.1  \\
        &                              & without & 12.5±3.8 & 10.2±1.2 & 10.1±1.8 & 7.0±2.2  \\
        & \multirow{2}{*}{\textit{GPT-4-Turbo}} & \cellcolor[HTML]{EFEFEF}with    & \cellcolor[HTML]{EFEFEF}9.6±3.2  & \cellcolor[HTML]{EFEFEF}10.0±1.6 & \cellcolor[HTML]{EFEFEF}9.2±2.1  & \cellcolor[HTML]{EFEFEF}5.0±1.2  \\
        &                              & without & 13.9±4.3 & 11.0±1.5 & 12.9±3.1 & 7.5±2.7  \\ 

        & \multirow{2}{*}{\textit{Claude-3.5-Sonnet}} & \cellcolor[HTML]{EFEFEF}with    & \cellcolor[HTML]{EFEFEF}\textbf{6.6±0.7} & \cellcolor[HTML]{EFEFEF}\textbf{8.0±0.0} & \cellcolor[HTML]{EFEFEF}\textbf{4.8±0.6} & \cellcolor[HTML]{EFEFEF}\textbf{4.4±0.9} \\
        &                                   & without & 11.9±3.4 & 8.1±0.3  & 5.8±1.1  & 5.9±1.4  \\
        \midrule

        \multirow{8}{*}{\textbf{5}} 
        & \multirow{2}{*}{\textit{Mixtral-7x8B}} & \cellcolor[HTML]{EFEFEF}with    & \cellcolor[HTML]{EFEFEF}10.6±1.4 & \cellcolor[HTML]{EFEFEF}13.1±2.7 & \cellcolor[HTML]{EFEFEF}13.0±2.1 & \cellcolor[HTML]{EFEFEF}9.7±2.5  \\
        &                               & without & 12.3±2.6 & 15.0±2.2 & 15.4±2.3 & 13.8±1.7 \\ 

        & \multirow{2}{*}{\textit{Llama3-70B}}  & \cellcolor[HTML]{EFEFEF}with    & \cellcolor[HTML]{EFEFEF}6.7±0.5  & \cellcolor[HTML]{EFEFEF}8.1±0.3  & \cellcolor[HTML]{EFEFEF}5.5±0.9  & \cellcolor[HTML]{EFEFEF}5.4±1.1  \\
        &                              & without & 8.4±1.7  & 9.3±0.5  & 8.7±1.2  & 5.4±1.4  \\
        & \multirow{2}{*}{\textit{GPT-4-Turbo}} & \cellcolor[HTML]{EFEFEF}with    & \cellcolor[HTML]{EFEFEF}6.8±0.6  & \cellcolor[HTML]{EFEFEF}8.4±0.5  & \cellcolor[HTML]{EFEFEF}6.6±1.3  & \cellcolor[HTML]{EFEFEF}4.6±0.9  \\
        &                              & without & 8.9±0.8  & 9.6±0.8  & 10.1±2.7 & 5.7±1.8  \\ 

        & \multirow{2}{*}{\textit{Claude-3.5-Sonnet}} & \cellcolor[HTML]{EFEFEF}with    & \cellcolor[HTML]{EFEFEF}\textbf{6.1±0.3} & \cellcolor[HTML]{EFEFEF}\textbf{8.0±0.0} & \cellcolor[HTML]{EFEFEF}\textbf{4.0±0.4} & \cellcolor[HTML]{EFEFEF}\textbf{3.0±0.0} \\
        &                                   & without & 7.5±0.9  & \textbf{8.0±0.0} & 4.9±0.5  & 4.0±1.0  \\
        \midrule

        \multirow{8}{*}{\textbf{20}} 
        & \multirow{2}{*}{\textit{Mixtral-7x8B}} & \cellcolor[HTML]{EFEFEF}with    & \cellcolor[HTML]{EFEFEF}10.1±1.2 & \cellcolor[HTML]{EFEFEF}11.2±2.3 & \cellcolor[HTML]{EFEFEF}10.7±1.2 & \cellcolor[HTML]{EFEFEF}8.8±2.0  \\
        &                               & without & 11.3±1.7 & 13.0±1.2 & 14.0±1.3 & 13.1±1.7 \\ 

        & \multirow{2}{*}{\textit{Llama3-70B}}  & \cellcolor[HTML]{EFEFEF}with    & \cellcolor[HTML]{EFEFEF}6.7±0.5  & \cellcolor[HTML]{EFEFEF}\textbf{8.0±0.0} & \cellcolor[HTML]{EFEFEF}4.7±0.5  & \cellcolor[HTML]{EFEFEF}5.2±1.2  \\
        &                              & without & 7.4±0.8  & 8.8±0.6  & 7.7±1.3  & 4.6±0.9  \\
        & \multirow{2}{*}{\textit{GPT-4-Turbo}} & \cellcolor[HTML]{EFEFEF}with    & \cellcolor[HTML]{EFEFEF}6.1±0.3  & \cellcolor[HTML]{EFEFEF}8.2±0.4  & \cellcolor[HTML]{EFEFEF}5.1±0.8  & \cellcolor[HTML]{EFEFEF}4.1±0.7  \\
        &                              & without & 7.3±0.6  & 9.1±0.3  & 7.7±1.8  & 5.0±1.5  \\ 

        & \multirow{2}{*}{\textit{Claude-3.5-Sonnet}} & \cellcolor[HTML]{EFEFEF}with    & \cellcolor[HTML]{EFEFEF}\textbf{6.0±0.0} & \cellcolor[HTML]{EFEFEF}\textbf{8.0±0.0} & \cellcolor[HTML]{EFEFEF}\textbf{3.6±0.5} & \cellcolor[HTML]{EFEFEF}\textbf{3.0±0.0} \\
        &                                   & without & 6.4±0.7  & \textbf{8.0±0.0} & 4.1±0.3  & 3.4±0.7  \\
        \bottomrule
    \end{tabular}
    \label{tab:table3}
\end{table}

To evaluate the effect of different backbone LLMs on our proposed method, we conducted an ablation study. In addition to \textit{Claude-3.5-Sonnet} \cite{anthropic2024claude}, we repeated the experiments in Table \ref{tab:table2} using three different backbones: two open-source LLMs (\textit{Mixtral-7x8B} \cite{jiang2024mixtral} and \textit{Llama3-70B} \cite{meta2024llama}) and one closed-source LLM (\textit{GPT-4-Turbo} \cite{openai2024gpt4turbo}), accessed via its API. We report the results for single-trial, 5-trial, and 20-trial runs, and also examine the impact of removing contextual domain knowledge from the prompts. As indicated by Table \ref{tab:table3}, \textit{Claude-3.5-Sonnet}, our choice, statistically outperforms the other LLMs across all four cases. \textit{Llama3-70B} follows as the second top performer in most cases, demonstrating that open-source LLMs can be effective alternatives. This suggests that our framework is compatible with different LLMs. We recommend users or researchers may start with \textit{Claude-3.5-Sonnet} for optimal performance, but may consider open-source models like \textit{Llama} for budget or customizability. Furthermore, in all cases, the results across all LLMs show that incorporating contextual domain knowledge consistently improves solution quality compared to the ones without knowledge. This finding aligns with our previous observation in Figure \ref{fig:fig1}, highlighting the robustness of our methods and the significant role of domain knowledge in enhancing LLMs' reasoning performance in such CO tasks with real-world contexts.

\section{Concluding Remarks}

Experimental results demonstrate that our proposed method, particularly when incorporating contextual domain knowledge, offer significant advantages in both convergence speed and solution quality compared to benchmarking stochastic and deterministic approaches. Across all experiments and backbone LLMs, the inclusion of domain knowledge consistently improved performance. While \textit{Claude-3.5-Sonnet} is the most effective backbone LLM, open-source models such as \textit{Llama3-70B} also shows strong performance. Overall, these findings illustrate the capability of our proposed method to effectively address DSM sequencing tasks, leveraging both semantic and mathematical reasoning of LLMs for real-world combinatorial optimization problems.

Despite the promising results, there are some limitations. First, the scope of the current experiments could be expanded by collecting more diverse DSM cases, including larger and more complex networks, to achieve more robust statistical results. Second, while we focused on DSM sequencing tasks, the effectiveness of our proposed method could be explored across a wider range of real-world combinatorial optimization tasks to assess its generalizability. Third, we plan to delve into the intermediate changes in LLM outputs during optimization in future work, which may provide greater interpretability and insights into the LLMs' reasoning processes.

In conclusion, our study highlights the potential of LLMs for combinatorial optimization of design structure matrix, showing that including domain knowledge can significantly enhance the performance of our methods. Future research can build on these findings to further refine LLM-driven methods and extend applications to more complex and diverse combinatorial optimization problems.

%\section*{Disclosure statement}
%No potential conflict of interest was reported by the author(s).

%\section*{Acknowledgements}
%This work was supported by the SUTD-MIT International Design Center, SUTD Data-Driven Innovation Laboratory (DDI, https://ddi.sutd.edu.sg/), and Shanghai Jiao Tong University under the grant of the National Natural Science Foundation of China (52035007, 51975360), Special Program for Innovation Method of the Ministry of Science and Technology, China (2018IM020100), and National Social Science Foundation of China (17ZDA020).

\bibliographystyle{unsrtnat}
\bibliography{references}

\section*{Appendix 1. Full prompts}

\subsection*{Prompt for our proposed method (the main method):}

\begin{tcolorbox}[colback=gray!10, colframe=white, coltitle=black, boxrule=0.5mm, arc=0mm, left=2mm, right=2mm, top=1mm, bottom=1mm, width=\textwidth]
You are an expert in the domain of combinational optimization.
\\
\\
Please assist me to find an optimal sequential order that minimizes feedback cycles in the dependency network described below. Your task is to propose a new order that differs from previous attempts and has fewer feedback cycles than any listed.
\\
\\
\texttt{<Description of the entire Network>} \texttt{\{network\_description\}} \texttt{</Description of the entire Network>}\\
\texttt{<Nodes with Descriptions>} \texttt{\{node\_list\_with\_description\}} \texttt{</Nodes with Descriptions>}\\
\texttt{<Edges>} \texttt{\{edge\_list\}} \texttt{</Edges>}
\\
\\
Below are some previous sequential orders arranged in descending order of feedback cycles (lower is better): \texttt{\{selected\_historical\_solutions\}}
\\
\\
Please suggest a new order that:\\
- Is different from all prior orders.\\
- Has fewer feedback cycles than any previous order.\\
- Covers all nodes exactly once.\\
- Starts with \texttt{<order>} and ends with \texttt{</order>}.\\
- You can use the descriptions of nodes and networks to support your suggestion.
\\
\\
Output Format:\\
\texttt{<order> ...... </order>}
\\
\\
Please provide only the order and nothing else.
\end{tcolorbox}

\subsection*{Prompt for our proposed method (removing contextual domain knowledge):}

\begin{tcolorbox}[colback=gray!10, colframe=white, coltitle=black, boxrule=0.5mm, arc=0mm, left=2mm, right=2mm, top=1mm, bottom=1mm, width=\textwidth]
You are an expert in the domain of combinational optimization.
\\
\\
Please assist me to find an optimal sequential order that minimizes feedback cycles in the dependency network described below. Your task is to propose a new order that differs from previous attempts and has fewer feedback cycles than any listed.
\\
\\
\texttt{<Nodes>} \texttt{\{node\_list\}} \texttt{</Nodes>}\\
\texttt{<Edges>} \texttt{\{edge\_list\}} \texttt{</Edges>}
\\
\\
Below are some previous sequential orders arranged in descending order of feedback cycles (lower is better): \texttt{\{selected\_historical\_solutions\}}
\\
\\
Please suggest a new order that:\\
- Is different from all prior orders.\\
- Has fewer feedback cycles than any previous order.\\
- Covers all nodes exactly once.\\
- Starts with \texttt{<order>} and ends with \texttt{</order>}.
\\
\\
Output Format:\\
\texttt{<order> ...... </order>}
\\
\\
Please provide only the order and nothing else.
\end{tcolorbox}

\subsection*{Example of \texttt{\{network\_description\}} in the UCAV design activity DSM case:}

\begin{tcolorbox}[colback=gray!10, colframe=white, coltitle=black, boxrule=0.5mm, arc=0mm, left=2mm, right=2mm, top=1mm, bottom=1mm, width=\textwidth]

\texttt{This network represents the dependency relationships among conceptual design activities for UCAV development at Boeing. Each node corresponds to a specific task or analysis, and directed edges indicate the prerequisite relationships between these tasks. Nodes: Each node is a task or analysis in the conceptual design process. Edges: Directed edges show the prerequisite relationships between these tasks and analyses.}

\end{tcolorbox}

\subsection*{Example of \texttt{\{node\_list\_with\_description\}} in the UCAV design activity DSM case:}

\begin{tcolorbox}[colback=gray!10, colframe=white, coltitle=black, boxrule=0.5mm, arc=0mm, left=2mm, right=2mm, top=1mm, bottom=1mm, width=\textwidth]

\texttt{[}\\
\texttt{\{'id': 'lzOtR', 'name': 'Create Configuration Concepts'\},}\\
\texttt{\{'id': 'yLlKi', 'name': 'Prepare UCAV Conceptual DR\&O'\},}\\
\texttt{\{'id': 'Swvi2', 'name': 'Prepare 3-View Drawing \& Geometry Data'\},}\\
\texttt{\{'id': 'CDcxF', 'name': 'Perform Weights Analyses \& Evaluation'\},}\\
\texttt{\{'id': '0KGDm', 'name': 'Perform Aerodynamics Analyses \& Evaluation'\},}\\
\texttt{\{'id': '4wHtv', 'name': 'Perform Multidisciplinary Analyses \& Evaluation'\},}\\
\texttt{\{'id': 'AgIBP', 'name': 'Prepare \& Distribute Choice Config. Data Set'\},}\\
\texttt{\{'id': 'gRtHi', 'name': 'Perform S\&C Characteristics Analyses \& Eval.'\},}\\
\texttt{\{'id': 'GV9RJ', 'name': 'Make Concept Assessment and Variant Decisions'\},}\\
\texttt{\{'id': 'I1j2m', 'name': 'Perform Performance Analyses \& Evaluation'\},}\\
\texttt{\{'id': 'Vzzm7', 'name': 'Perform Propulsion Analyses \& Evaluation'\},}\\
\texttt{\{'id': 'B0BFG', 'name': 'Perform Mechanical \& Electrical Analyses \& Eval.'\}}\\
\texttt{]}
\end{tcolorbox}

\subsection*{Example of \texttt{\{node\_list\}} in the UCAV design activity DSM case:}

\begin{tcolorbox}[colback=gray!10, colframe=white, coltitle=black, boxrule=0.5mm, arc=0mm, left=2mm, right=2mm, top=1mm, bottom=1mm, width=\textwidth]

\texttt{['lzOtR', 'yLlKi', 'Swvi2', 'CDcxF', '0KGDm', '4wHtv', 'AgIBP', 'gRtHi', 'GV9RJ', 'I1j2m', 'Vzzm7', 'B0BFG']}

\end{tcolorbox}

\subsection*{Example of \texttt{\{edge\_list\}} in the UCAV design activity DSM case:}

\begin{tcolorbox}[colback=gray!10, colframe=white, coltitle=black, boxrule=0.5mm, arc=0mm, left=2mm, right=2mm, top=1mm, bottom=1mm, width=\textwidth]

\texttt{[}\\
\texttt{\{'dependent': '0KGDm', 'predecessor': 'Swvi2'\},}\\
\texttt{\{'dependent': 'AgIBP', 'predecessor': 'lzOtR'\},}\\
\texttt{\{'dependent': '0KGDm', 'predecessor': 'yLlKi'\},}\\
\texttt{\{'dependent': 'Swvi2', 'predecessor': 'lzOtR'\},}\\
\texttt{\ldots}\\
\texttt{]}

\end{tcolorbox}

\subsection*{Example of \texttt{\{selected\_historical\_solutions\}} in the UCAV design activity DSM case:}

\begin{tcolorbox}[colback=gray!10, colframe=white, coltitle=black, boxrule=0.5mm, arc=0mm, left=2mm, right=2mm, top=1mm, bottom=1mm, width=\textwidth]

\texttt{[}\\
\texttt{\{'solution': 'lzOtR, yLlKi, GV9RJ, AgIBP, B0BFG, Vzzm7, Swvi2, CDcxF, 0KGDm, I1j2m, gRtHi, 4wHtv', 'score': 15.0\},}\\
\texttt{\{'solution': 'B0BFG, yLlKi, Vzzm7, lzOtR, Swvi2, CDcxF, AgIBP, 0KGDm, GV9RJ, I1j2m, gRtHi, 4wHtv', 'score': 13.0\},}\\
\texttt{\ldots}\\
\texttt{]}

\end{tcolorbox}

\section*{Appendix 2. Parameter settings for three variants of GA}

In all three variants of the GA used for DSM sequencing tasks, we employed the following shared settings: the number of generations was set to 2,000, with a selection mechanism using tournament selection and mutation using shuffled indexes. The GA was implemented using the DEAP library with standard configurations for initial population generation \cite{fortin2012deap}. The different configurations for each GA setting are shown in the table below. The meaning of each parameter also refers to \cite{fortin2012deap}.

\begin{table}[h!]
    \centering
    \renewcommand{\arraystretch}{1.3} % Adjust row height
    \setlength{\tabcolsep}{8pt} % Adjust column spacing
    \begin{tabular}{p{3.5cm} m{1.4cm}<{\centering} m{3cm}<{\centering} m{1.75cm}<{\centering} m{1.75cm}<{\centering} m{1.75cm}<{\centering}}
        \toprule
        & \textbf{Population} & \textbf{Individual Mutation Probability} & \textbf{Tournament Size} & \textbf{Crossover Probability} & \textbf{Mutation Probability} \\
        \midrule
        \textbf{Exploration-focused} & 50 & 0.05 & 5 & 0.6 & 0.4 \\
        \textbf{Exploitation-focused} & 10 & 0.01 & 20 & 0.9 & 0.1 \\
        \textbf{Balanced} & 20 & 0.02 & 10 & 0.7 & 0.3 \\
        \bottomrule
    \end{tabular}
    \label{tab:parameter_settings}
\end{table}

\end{document}